\documentclass[sigconf,screen,nonacm]{acmart}

\setcopyright{none}
\renewcommand\footnotetextcopyrightpermission[1]{}
\def\BibTeX{{\rm B\kern-.05em{\sc i\kern-.025em b}\kern-.08emT\kern-.1667em\lower.7ex\hbox{E}\kern-.125emX}}

\usepackage{nicefrac}
\usepackage{siunitx}
\usepackage{array,framed}
\usepackage{booktabs}
\usepackage{caption}

\usepackage{
  color,
  float,
  epsfig,
  wrapfig,
  graphics,
  graphicx,
  subcaption
}

\usepackage{textcomp,amssymb}
\usepackage{setspace}
\usepackage{latexsym,url}
\usepackage{enumerate}
\usepackage{algorithm2e}
\usepackage{algpseudocode}
\usepackage{xparse} 
\usepackage{xspace}
\usepackage{multirow}
\usepackage{csvsimple}
\usepackage{balance}

\usepackage{
  tikz,
  pgfplots,
  pgfplotstable
}
\usepackage{hyperref}

\usetikzlibrary{
  shapes.geometric,
  arrows,
  external,
  pgfplots.groupplots,
  matrix
}

\pgfplotsset{compat=1.9}

\usepackage{mathtools}

\DeclareMathAlphabet{\mathcal}{OMS}{cmsy}{m}{n}

\DeclareGraphicsExtensions{%
    .png,.PNG,%
    .pdf,.PDF,%
    .jpg,.mps,.jpeg,.jbig2,.jb2,.JPG,.JPEG,.JBIG2,.JB2}

\usepackage[]{todonotes}
\usepackage{amsmath}
\usepackage{amssymb}

\begin{document}

\title{Temporal Poisoning:\\Clean-Label Backdoors via Event Redistribution in SNNs}

\author{Roberto Riaño}
\affiliation{%
  \institution{Radboud University, The Netherlands}
}
\affiliation{%
  \institution{Ikerlan Research Centre, Spain}
}
\email{roberto.rianohidalgo@ru.nl}

\author{Gorka Abad}
\affiliation{%
  \institution{University of Bergen, Norway}
}
\email{gorka.abad@uib.no}

\author{Stjepan Picek}
\affiliation{%
  \institution{University of Zagreb Faculty of Electrical Engineering and Computing, Croatia}
}
\affiliation{%
\institution{Radboud University, The Netherlands}
}
\email{stjepan.picek@ru.nl}

\author{Aitor Urbieta}
\affiliation{%
  \institution{Ikerlan Research Centre, Spain}
}
\email{aurbieta@ikerlan.es}

\date{}

\begin{abstract}
Backdoor attacks on Spiking Neural Networks (SNNs) have primarily assumed dirty-label poisoning, in which triggered training samples are relabeled to an attacker-selected class. We study clean-label temporal poisoning, where a fixed timestamp transformation is applied only to the target-class training streams, leaving their labels unchanged. The transformation preserves the per-pixel, per-polarity event count exactly, making clean and triggered samples identical after temporal aggregation while altering the sequence processed by the SNN. Across three neuromorphic datasets and both convolutional and transformer-based victims, the attack reaches an ASR of $1.00$ in strongest configurations. We analyze the attack through poison-budget and trigger-shape ablations and evaluate established backdoor defenses adapted to spiking models. Defenses that collapse the time axis before inspection are blind by construction, while feature-space methods detect the poison only in selected settings. Our model-free detector, based on per-step event mass, detects the evaluated temporal transformations, demonstrating both the limitation of rate-collapsed defenses and the boundary of the attack's stealth. To our knowledge, this is the first clean-label backdoor attack evaluated on SNNs and neuromorphic event data.
\end{abstract}

\maketitle

\keywords{Spiking neural networks, Backdoor attack, Clean-label}

\section{Introduction}
\label{sec:intro}

Deep Neural Networks (DNNs) have achieved remarkable performance across a wide range of tasks, from image recognition~\cite{krizhevsky2012imagenet} to natural language processing~\cite{vaswani2017attention} and speech recognition~\cite{hinton2012speech}. However, this performance also carries the need for large models that demand substantial computational resources and energy for both training and deployment~\cite{strubell2019energy}. These costs have motivated the search for more efficient alternatives, such as Spiking Neural Networks (SNNs), which have emerged as a biologically plausible and energy-efficient option~\cite {tavanaei2019deep}.

SNNs differ from traditional DNNs in how they represent and process information. Instead of operating on continuous activations, SNNs encode information as sequences of impulses, or spikes, over time. Unlike DNNs, which activate all their neurons for every input, SNNs integrate incoming spikes into a membrane potential and fire only when the potential exceeds a threshold, remaining inactive otherwise. This event-driven behavior enables SNNs to process sparse inputs with very low power consumption~\cite{tavanaei2019deep,davies2018loihi}, and it naturally captures the temporal structure of the data, where the precise timing of spikes carries the information rather than their aggregate count alone. This property makes SNNs particularly effective when paired with neuromorphic data captured by Dynamic Vision Sensor (DVS) cameras, which asynchronously record per-pixel brightness changes at high temporal resolution and low latency~\cite{lichtsteiner2008dvs}, instead of full frames captured by conventional cameras at a fixed frame rate.

The combination of energy efficiency and low latency has made SNNs attractive for edge AI systems and neuromorphic hardware deployments~\cite{davies2018loihi}. They are increasingly being considered for safety-critical applications such as autonomous driving~\cite{chen2020eventbased,viale2021carsnn} and robotics~\cite{pfeiffer2013gridbot,loihi2020trajectory}, where dedicated neuromorphic processors are already being integrated. As SNNs move toward these settings, understanding their vulnerabilities becomes essential for them to be trusted in real deployments.

Backdoor attacks are one of the main threats to these systems. In a backdoor attack, an adversary manipulates the training process so that the resulting model behaves normally under clean inputs, but produces an attacker-chosen output whenever a specific trigger is present~\cite{gu2019badnets}. Recent work has shown that SNNs are vulnerable to such attacks in both digital and physical scenarios. Digital attacks embed spatio-temporal triggers directly into the event data~\cite{abad2022poster,abad2024sneaky}, while physical attacks reproduce the trigger in the real world through the DVS sensor~\cite{physicalsnn}. In all of these studies, the poisoned samples are mislabeled, i.e., their labels are changed to the target class so that the model associates the trigger with that class. These are known as dirty-label backdoor attacks and have been shown to be highly effective on neuromorphic systems, but this label modification introduces a weakness. Since the content of a poisoned sample no longer matches its label, the poison can be exposed by human inspection or removed with a relabeling pass over the training data.



Clean-label attacks avoid explicit label inconsistency by leaving the assigned label of every poisoned sample unchanged. A common formulation modifies only samples that already belong to the target class, preventing a defender from identifying the poison through label inspection or a simple relabeling pass. Clean-label backdoors have been extensively studied in the image domain~\cite{turner2019label,barni2019new,saha2020hidden,zhao2020cleanlabel,souri2022sleeper}, where they are regarded as a stronger data-poisoning threat than dirty-label attacks. To our knowledge, clean-label data poisoning has not previously been evaluated for SNNs. This gap is important because neuromorphic data exposes a temporal attack surface: a trigger can be encoded in the timing of events rather than their location, leaving an aggregate spatial representation unchanged while modifying the dynamics processed by a spiking model.

In this work, we introduce a clean-label temporal backdoor for SNNs. The attack remaps event timestamps while preserving the exact per-pixel, per-polarity event count for each sample. Consequently, the clean and triggered rate frames are identical, although their per-step event distributions and the resulting spiking trajectories differ. We evaluate three temporal transformations on convolutional and transformer-based SNNs and study both established backdoor defenses and an adaptive detector that directly examines temporal event mass.

Our main contributions are as follows:
\begin{itemize}
    \item We introduce the first clean-label temporal poisoning attack on SNNs that preserves the per-pixel, per-polarity event count exactly, yielding identical clean and triggered rate frames.
    \item We evaluate three timestamp-remapping families across three neuromorphic datasets and two SNN architectures, characterizing attack effectiveness, clean-accuracy cost, poison budget, and sensitivity to trigger shape and placement.
    \item We show that the representation exposed to a defense is decisive: methods that collapse time before inspection discard the trigger signal, whereas time-aware analysis can expose it. We demonstrate this with a model-free detector based on per-step event mass.
\end{itemize}
\section{Background}
\label{sec:background}




Unlike a conventional frame-based camera, which captures full images at a fixed rate, a neuromorphic sensor reacts to per-pixel brightness changes and reports each one as a separate event. Its output is therefore an asynchronous stream of events rather than a sequence of frames. We write such a stream as $\mathcal{E} = \{e_i\}_{i=1}^{N}$, where each event $e_i = (x_i, y_i, t_i, p_i)$ records the pixel location $(x_i, y_i)$ at which the change occurred, the time $t_i \in [0, \Delta]$ at which it occurred within a recording of duration $\Delta$, and a polarity $p_i \in \{-1, +1\}$ indicating whether the brightness increased or decreased.

To train a spiking network, the events are first discretized into a fixed number of consecutive time windows, which turns the continuous stream into a sequence of tensors that the network reads. The binning operator $\mathrm{B}$ splits the stream into $T$ time bins and produces a tensor $\mathbf{X} \in \mathbb{N}^{\,T \times 2 \times H \times W}$, where $\mathbf{X}[\tau, c, y, x]$ counts the events at pixel $(x, y)$ with polarity channel $c$ whose timestamp falls in bin $\tau$. We write $\mathbf{X}[\tau] \in \mathbb{N}^{2 \times H \times W}$ for the slice at bin $\tau$.

Collapsing this sequence over time yields a single map that records, for each pixel and polarity, how many events occurred in total. We call this the \emph{rate frame}, and define it as the sum of the binned tensor over time,
\begin{equation}
  \mathbf{R}(\mathcal{E}) \;=\; \sum_{\tau=1}^{T} \mathbf{X}[\tau]
  \;\in\; \mathbb{N}^{\,2 \times H \times W}.
  \label{eq:rateframe}
\end{equation}
The rate frame is the representation seen by any inspection that collapses the time axis. It is what a spatial visual check, and the spatial or rate-based defenses we later evaluate, operate on. Our attack is designed so that this view carries no evidence of the trigger.

\subsection{Spiking Dynamics}
\label{sec:snn}

The victim is a spiking neural network trained in activation-based multi-step mode, built from leaky integrate-and-fire (LIF) neurons. Intuitively, a LIF neuron accumulates its incoming inputs into an internal membrane potential that leaks toward zero over time. When this potential crosses a threshold, the neuron emits a spike and its potential is reset; otherwise it stays silent (emits no output spike), and the potential carries over to the next step. Formally, for a neuron with input current $X[\tau]$ at bin $\tau$, the membrane potential evolves as
\begin{align}
  H[\tau] &= \beta\, V[\tau-1] + X[\tau], \label{eq:lif-charge}\\
  S[\tau] &= \Theta\!\big(H[\tau] - V_{\mathrm{th}}\big), \label{eq:lif-fire}\\
  V[\tau] &= H[\tau]\,(1 - S[\tau]) + V_{\mathrm{reset}}\, S[\tau],
  \label{eq:lif-reset}
\end{align}
where $\beta \in (0,1)$ is the leak factor, $V_{\mathrm{th}}$ is the firing threshold, $\Theta$ is the Heaviside step, and $S[\tau] \in \{0,1\}$ is the output spike.

The firing step in \eqref{eq:lif-fire} is non-differentiable. Thus, in order to train the network, we replace this step in the backward pass with a smooth approximation, the ATan surrogate gradient,
\begin{equation}
  \frac{\partial S[\tau]}{\partial H[\tau]}
  \;\approx\;
  \frac{\alpha}{2}\,
  \frac{1}{1 + \left(\tfrac{\pi}{2}\,\alpha\,(H[\tau] - V_{\mathrm{th}})\right)^{2}},
  \label{eq:atan}
\end{equation}
with slope parameter $\alpha$. Finally, class scores are produced by a voting readout that aggregates the output-population spikes over all $T$ steps.

The key property for this work is that \eqref{eq:lif-charge} carries state across bins through the term $\beta\, V[\tau-1]$. The spike pattern, therefore, depends on when inputs arrive and on how they coincide, not only on their total count. A rate frame reports the total count and discards this ordering, so two streams with identical rate frames can drive a spiking network to very different internal trajectories. This gap is what the temporal trigger exploits.
\section{Threat Model}
\label{sec:threat}

We consider a clean-label data-poisoning attacker that operates entirely in the digital domain. Unlike physical backdoors that reproduce a trigger through the sensor at capture time~\cite{physicalsnn}, the attacker does not modify the camera or the observed scene; it can only tamper with the event streams supplied to the victim for training. This setting applies when the victim relies on a public or third-party dataset, accepts externally contributed training data, or outsources part of training to an MLaaS provider. Concretely, our evaluation instantiates an update-stage poisoning scenario, in which an initially clean model is subsequently trained on data containing temporally transformed target-class samples.

\paragraph{Capabilities.}
The attacker can replace a bounded fraction of training samples from a chosen target class $y_t$ with timestamp-remapped versions of those same samples. It cannot change their labels, modify the model architecture, optimizer, loss, training code, or trained weights, and it has no query access to the victim during poisoning. The attack is therefore confined to the input stream of target-class samples and does not manipulate neurons directly. As the poison lives in the data rather than the training schedule, the poisoned samples may appear at any stage of training, including pretraining.

\paragraph{Clean-label constraint.}
Every poisoned sample retains its original label $y_t$, and the attacker modifies only samples drawn from that class. Thus, label inspection and a relabeling pass cannot identify or remove the poison. Our stealth claim is representation-specific: the transformation preserves the rate frame exactly, but it can remain visible to a defender that inspects the full temporal event sequence.

\paragraph{Knowledge.}
The attacker knows the event representation and temporal discretization used by the victim, which is sufficient to construct the trigger. Trigger parameters are fixed before the corresponding victim is trained, and the evaluation reports all three trigger families rather than assuming an oracle that selects the best family after observing victim behavior.

\paragraph{Goal.}
At test time, applying the trigger to a non-target input should cause the prediction of $y_t$. At the same time, the poisoned victim should retain useful accuracy on clean, untriggered inputs. We therefore evaluate attack success jointly with the change in clean accuracy relative to a clean model trained under the same epoch budget.
\section{Method}
\label{sec:method}

We describe a clean-label backdoor whose trigger is encoded in the timing of neuromorphic events rather than in their spatial location. The attack changes only when events occur. It therefore preserves the time-collapsed rate frame exactly, while altering the temporal sequence processed by the spiking network.

\subsection{Temporal Trigger}
\label{sec:trigger}

A temporal trigger is a fixed rule that maps each event to a new point in time while leaving its pixel location and polarity unchanged. Concretely, let $\varphi:[0,\Delta]\rightarrow[0,\Delta]$ be a timestamp-remapping function defined over a recording interval of duration $\Delta$. We apply the same transformation to every event in a stream:
\begin{equation}
  \varphi(\mathcal{E})
  =
  \big\{\,(x_i,\,y_i,\,\varphi(t_i),\,p_i)\,\big\}_{i=1}^{N}.
  \label{eq:trigger-apply}
\end{equation}
Only the timestamp $t_i$ is replaced. The trigger never adds, removes, changes the polarity of, or spatially relocates an event.

We instantiate $\varphi$ using three remapping families that differ in how strongly they modify the temporal coincidence structure of the event stream.

\paragraph{Concentrate.}
All events are compressed into a narrow temporal window of width
$w\ll\Delta$ beginning at $t_0$:
\begin{equation}
  \varphi_{\mathrm{conc}}(t)
  =
  t_0+\frac{w}{\Delta}t.
  \label{eq:conc}
\end{equation}
This transformation packs the stream into a short burst, causing many events that were previously separated in time to arrive within a small number of bins.

\paragraph{Front-load.}
Events are moved toward the beginning of the recording using a power map with exponent $\gamma>1$:
\begin{equation}
  \varphi_{\mathrm{front}}(t)
  =
  \Delta\left(\frac{t}{\Delta}\right)^{\gamma}.
  \label{eq:front}
\end{equation}
Because $\gamma>1$, $\varphi_{\mathrm{front}}(t)\leq t$ for $t\in[0,\Delta]$. The event order is preserved, but a larger fraction of the event mass is assigned to earlier time steps.

\paragraph{Shift.}
A constant temporal offset $\delta$ is applied with wraparound:
\begin{equation}
  \varphi_{\mathrm{shift}}(t)
  =
  (t+\delta)\bmod\Delta.
  \label{eq:shift}
\end{equation}
This transformation preserves the inter-event intervals except at the recording boundary and primarily changes the absolute phase of the stream. We include it as a minimal-perturbation baseline for testing whether a phase change alone is sufficient to install a backdoor.

In practice, these transformations act on the $T$ discrete bins, and we report each family by the parameter swept in Section~\ref{sec:evaluation} and Figure~\ref{fig:triggershape}. For concentrate this is the target bin $t^{*}$ at which the burst is placed, so $t_0 = t^{*}\,\Delta/T$ and $w$ spans a single bin. For shift it is the roll amount $s$ in bins, so $\delta = s\,\Delta/T$; equivalently, the shift rolls the frame axis, $\mathbf{X}[\tau]\mapsto\mathbf{X}\!\big[((\tau-1+s)\bmod T)+1\big]$. For front-loading, it is the number of leading bins $k$ into which the event mass is drawn.

\subsection{Clean-Label Poisoning}
\label{sec:poison}

The attacker selects a target class $y_t$ and a poison rate $\rho\in(0,1)$. A fraction $\rho$ of the target-class training streams is replaced by their temporally transformed versions. Their labels remain unchanged, so every poisoned sample retains its original target-class label $y_t$.

Let $\mathcal{D}_p$ denote the resulting poisoned training set. The victim is trained normally by minimizing
\begin{equation}
  \min_{\theta}
  \mathbb{E}_{(\mathcal{E},y)\sim\mathcal{D}_p}
  \left[
    \mathcal{L}\big(f_\theta(\mathcal{E}),y\big)
  \right],
  \label{eq:train}
\end{equation}
where $f_\theta$ is the spiking victim and $\mathcal{L}$ is the classification loss. No backdoor-specific loss, model modification, or weight manipulation is required. For simplicity, we write $f_\theta(\mathcal{E})$ for the network applied to the binned stream $\mathrm{B}(\mathcal{E})$.

The attack seeks a model satisfying
\begin{equation}
  f_\theta\big(\varphi(\mathcal{E})\big)=y_t
  \text{ for inputs with }y\neq y_t,\quad
  f_\theta(\mathcal{E})=y
  \text{ on clean inputs.}
  \label{eq:goal}
\end{equation}

The first condition measures attack effectiveness: applying the fixed temporal transformation to a non-target input results in the prediction of the attacker-selected class. The second captures clean utility. In practice, we jointly evaluate the two objectives using attack success rate and clean accuracy, since a successful attack may still incur a detectable utility cost in some dataset-victim configurations.

\section{Attack Evaluation}
\label{sec:evaluation}
 
Every raw event stream is normalized to a common recording interval and discretized into $T=16$ consecutive bins of equal temporal duration, while preserving the two polarity channels. Each bin records the number of events occurring at every pixel and polarity during its corresponding time interval. A sample is therefore represented as a tensor of shape $[T,2,H,W]$, which the network uses in SpikingJelly~\cite{spikingjelly} activation-based multi-step mode. The network state is reset between batches, and the class logits are obtained by averaging the per-step logits over the $T$ time steps.
 
We evaluate on three neuromorphic datasets: N-MNIST~\cite{orchard2015converting} ($34\times34$, 60k train / 10k test, 10 classes), DVS-Gesture~\cite{amir2017low} ($128\times128$, 1176 train / 288 test, 11 classes), and CIFAR10-DVS~\cite{li2017cifar10} ($128\times128$, 9k train / 1k test, 10 classes). For each dataset, we train two victims of different families. The first is a convolutional SNN (conv-SNN) built from stacked $\text{Conv}\to\text{BN}\to\text{spiking neuron}\to\text{MaxPool}$ blocks followed by two linear layers and a voting layer. The second is SpikformerLite~\cite{li2024spikeformer}, a spiking transformer with dimension 128, depth 2, 8 heads, and 0.48\,M parameters. The conv-SNN uses IF neurons on N-MNIST (17.1\,M parameters) and LIF neurons with $\tau = 2.0$ on DVS-Gesture (1.70\,M) and CIFAR10-DVS (4.69\,M); across both families every neuron uses $V_\mathrm{th} = 1.0$, $V_\mathrm{reset} = 0.0$, hard reset, detached reset, and ATan surrogate gradients with slope $\alpha = 2.0$.
 
All victims are trained with Adam at learning rate $10^{-3}$, a cosine-annealing schedule, and cross-entropy on the time-averaged logits. Training runs a clean pretrain followed by a poisoning phase: N-MNIST uses 5 pretrain and 12 poison epochs at batch 64, DVS-Gesture uses 54 and 10 at batch 32, and CIFAR10-DVS uses 18 and 10 at batch 32. Every configuration is trained with three seeds, and the clean baseline for each dataset and architecture is trained on clean data only under the same total epoch budget.
 
For every dataset, victim, and run, we fix the target class to $y_t = 0$ and measure ASR over all non-target test samples. We also tested other target classes, and the attack works similarly; we exclude the results due to the page limit and to improve readability. We poison a fraction $\rho = 0.1$ of the target-class samples on N-MNIST and CIFAR10-DVS, and $\rho = 0.5$ on DVS-Gesture, whose target-class pool is far smaller; the effect of other poison rates is studied in Section~\ref{sec:exp-ablation}. The trigger instances are concentrated at frame $t^\ast = 8$ on N-MNIST and CIFAR10-DVS and at $t^\ast = 0$ on DVS-Gesture, front-loaded with $k = 4$, and shifted with $s = 8$. Since only target-class samples are poisoned, the absolute number of poisoned samples depends on the size of that pool, as reported in Table~\ref{tab:poolsize}. For comparability across datasets and defenses, the main tables use
$y_t=0$. We additionally repeated the attack-effectiveness evaluation
with other target classes, confirming the attack is not conditional on a specific class. These experiments are available in our repository. 
 
\begin{table}[t]
\centering
\caption{Number of poisoned samples at each poison rate $\rho$, given the size of the target-class pool ($y_t = 0$). Counts are $\max(1, \lfloor \rho\,|\mathcal{D}_{y_t}| \rfloor)$. Bold marks the operating point used in Table~\ref{tab:effectiveness}.}
\label{tab:poolsize}
\resizebox{\columnwidth}{!}{%
\begin{tabular}{lrrrrrrrr}
\toprule
 & & \multicolumn{7}{c}{Poison rate $\rho$} \\
\cmidrule(lr){3-9}
Dataset & Pool & 0.01 & 0.05 & 0.1 & 0.2 & 0.3 & 0.4 & 0.5 \\
\midrule
N-MNIST     & 5923 & 59 & 296 & \textbf{592} & 1184 & 1776 & 2369 & 2961 \\
CIFAR10-DVS &  900 &  9 &  45 &  \textbf{90} &  180 &  270 &  360 &  450 \\
DVS-Gesture &   97 &  1 &   4 &           9 &   19 &   29 &   38 & \textbf{48} \\
\bottomrule
\end{tabular}
}
\end{table}
 
We report clean accuracy (CA) on untriggered test inputs and attack success rate (ASR) on triggered non-target test inputs. ASR is the fraction of non-target samples classified as the attacker-selected class $y_t$ after applying the temporal transformation. We assess representation-specific stealth using the SSIM and $L_p$ distances between the clean and triggered rate frames. All reported values are the mean $\pm$ standard error over three random seeds.
 
\subsection{Attack Effectiveness}
\label{sec:exp-effectiveness}
 
Table~\ref{tab:effectiveness} reports the effectiveness and clean utility of the three temporal triggers. At least one trigger reaches an ASR of $1.00$ in five of the six dataset--victim pairs. The utility cost, however, differs substantially across settings. On N-MNIST, the poisoned victims remain within two percentage points of their clean baselines. The DVS-Gesture convolutional victim similarly retains its baseline accuracy. In contrast, the DVS-Gesture SpikformerLite victim and both CIFAR10-DVS victims experience larger clean-accuracy losses. The results, therefore, demonstrate the feasibility of temporal clean-label poisoning, but not uniformly utility-preserving behavior across all configurations.

We also test whether the timestamp transformations themselves cause a clean, unpoisoned model to predict the target class. As shown in Table~\ref{tab:cleanmodel}, applying the triggers to non-target inputs of a clean N-MNIST convolutional victim produces the target class in at most $1.18\%$ of cases. This is orders of magnitude below the ASR obtained after poisoning and shows that the high target prediction rate is learned during poisoned training rather than being an inherent bias of the transformation.

No trigger dominates across all victims. Concentrate reaches an ASR of $1.00$ on both N-MNIST victims and on the DVS-Gesture SpikformerLite victim, while front-load reaches $1.00$ on both CIFAR10-DVS victims. The trigger-shape ablation in Section~\ref{sec:exp-ablation} further shows that concentrate is sensitive to burst placement. In particular, its middle-window configuration is weaker on CIFAR10-DVS than configurations placed near either temporal boundary. These results show that temporal backdoor effectiveness depends on both the victim and the temporal transformation. We report all pre-specified trigger families rather than assuming that an attacker can select the best one using feedback from the victim.

Shift is consistently the weakest trigger. Its ASR remains near zero in almost every configuration, with the largest value being $0.22$ on the N-MNIST convolutional victim. Unlike concentrate and front-load, circular shifting largely preserves the local coincidence structure of the stream and changes primarily its absolute temporal phase. Its failure, therefore, provides evidence that learning the backdoor depends on a repeatable change in the distribution and coincidence of events, rather than on an arbitrary timestamp modification. We examine this interpretation further in Section~\ref{sec:mechanism}.
 

\begin{table}[t]
\centering
\footnotesize
\caption{Temporal transformations applied to a clean, unpoisoned
N-MNIST convolutional victim. The second column reports the percentage
of triggered non-target samples assigned to the target class by the
clean model. The third column reports ASR after poisoning under the
corresponding trigger. The transformations alone do not produce a
substantial target-class bias. Mean $\pm$ standard error over three
seeds.}
\label{tab:cleanmodel}
\begin{tabular}{lcc}
\toprule
Input &
\begin{tabular}[c]{@{}c@{}}Target prediction rate,\\clean model (\%)\end{tabular} &
\begin{tabular}[c]{@{}c@{}}ASR,\\poisoned model\end{tabular} \\
\midrule
No trigger   & $0.10 \pm 0.01$ & ---  \\
concentrate  & $0.00 \pm 0.00$ & 1.00 \\
front-load   & $0.53 \pm 0.41$ & 0.69 \\
shift        & $1.18 \pm 0.22$ & 0.22 \\
\bottomrule
\end{tabular}
\end{table}

\begin{table}[t]
\centering
\footnotesize
\setlength{\tabcolsep}{3.5pt}
\caption{Attack effectiveness and clean utility. Clean accuracy (CA) is measured on untriggered test inputs, and attack success rate (ASR) on triggered non-target test inputs. Results are mean $\pm$ standard error over three seeds. The target class is fixed to $y_t=0$. The poison rate is $\rho=0.1$ on N-MNIST and CIFAR10-DVS and $\rho=0.5$ on DVS-Gesture. Concentrate uses $t^\ast=8$ except on DVS-Gesture, where $t^\ast=0$.}
\label{tab:effectiveness}
\begin{tabular}{lcccc}
\toprule
 & \multicolumn{2}{c}{conv-SNN} & \multicolumn{2}{c}{SpikformerLite} \\
\cmidrule(lr){2-3}\cmidrule(lr){4-5}
Trigger & CA & ASR & CA & ASR \\
\midrule
\multicolumn{5}{l}{N-MNIST}\\
clean       & $0.99 \pm 0.00$ & --              & $0.99 \pm 0.00$ & --              \\
concentrate & $0.99 \pm 0.00$ & $1.00 \pm 0.00$ & $0.97 \pm 0.01$ & $1.00 \pm 0.00$ \\
front-load  & $0.99 \pm 0.00$ & $0.69 \pm 0.02$ & $0.98 \pm 0.00$ & $0.99 \pm 0.01$ \\
shift       & $0.99 \pm 0.00$ & $0.22 \pm 0.07$ & $0.98 \pm 0.00$ & $0.00 \pm 0.00$ \\
\midrule
\multicolumn{5}{l}{DVS-Gesture}\\
clean       & $0.78 \pm 0.01$ & --              & $0.93 \pm 0.00$ & --              \\
concentrate & $0.77 \pm 0.01$ & $0.84 \pm 0.09$ & $0.70 \pm 0.01$ & $1.00 \pm 0.00$ \\
front-load  & $0.77 \pm 0.01$ & $0.73 \pm 0.11$ & $0.67 \pm 0.03$ & $0.53 \pm 0.11$ \\
shift       & $0.77 \pm 0.00$ & $0.02 \pm 0.01$ & $0.68 \pm 0.05$ & $0.03 \pm 0.02$ \\
\midrule
\multicolumn{5}{l}{CIFAR10-DVS}\\
clean       & $0.62 \pm 0.01$ & --              & $0.60 \pm 0.01$ & --              \\
concentrate & $0.54 \pm 0.01$ & $0.78 \pm 0.22$ & $0.49 \pm 0.02$ & $0.63 \pm 0.31$ \\
front-load  & $0.54 \pm 0.02$ & $1.00 \pm 0.00$ & $0.46 \pm 0.01$ & $1.00 \pm 0.00$ \\
shift       & $0.55 \pm 0.01$ & $0.05 \pm 0.01$ & $0.48 \pm 0.03$ & $0.04 \pm 0.02$ \\
\bottomrule
\end{tabular}%
\end{table}
 
\subsection{Stealth}
\label{sec:exp-stealth}
 
The temporal transformations provide exact invariance under time-axis aggregation. For every evaluated sample and trigger, the SSIM between the clean and triggered rate frames is $1.00$, while the corresponding $L_0$ and $L_\infty$ distances are both zero. Thus, neither the pixel nor the polarity count changes after summing over time. Any manual inspection or defense that receives only the rate frame is therefore given an input that is mathematically identical to its clean counterpart.

This property should not be interpreted as universal perceptual invisibility. The temporal sequence itself changes substantially. On N-MNIST, the average SSIM between corresponding individual frames is $0.24$ for concentrate, $0.26$ for front-load, and $0.20$ for shift. A defender that preserves and directly examines the time axis can therefore observe the redistribution. The attack is stealthy with respect to rate-collapsed representations, not with respect to the full event sequence.

Table~\ref{tab:stealth} compares this rate-frame property with several spatial clean-label triggers on the N-MNIST convolutional victim. The additive and random-flip triggers reach ASR $1.00$, but visibly modify the rate frame, with SSIM values of $0.22$ and $0.31$. The corner-patch and universal-perturbation baselines preserve more of the rate frame, but do not establish a successful backdoor in this configuration. The temporal trigger is the only evaluated trigger that combines ASR $1.00$ with exact rate-frame equality. Because the spatial baselines were selected from exploratory single-seed sweeps, this comparison is descriptive rather than a statistically controlled ranking of trigger families.
 
\begin{table}[t]
\centering
\footnotesize
\setlength{\tabcolsep}{4pt}
\caption{Rate-frame invariance compared with spatial clean-label triggers on the N-MNIST convolutional victim. SSIM is measured after summing over time, so $1.00$ indicates exact equality of the time-collapsed representation. Spatial-trigger results are the best values observed in exploratory single-seed sweeps and are included as a descriptive comparison.}
\label{tab:stealth}
\begin{tabular}{@{}lccccc@{}}
\toprule
     & Temporal & Additive & Random- & Patch & Universal \\
     & (ours)   & (AE)     & flip    &       &           \\
\midrule
ASR  & $1.00$ & $1.00$ & $1.00$ & $0.19$ & $0.00$ \\
SSIM & $1.00$ & $0.22$ & $0.31$ & $0.74$ & $0.76$ \\
\bottomrule
\end{tabular}
\end{table}
 
\subsection{Ablation Studies}
\label{sec:exp-ablation}
 
\begin{figure}[t]
  \centering
  \includegraphics[width=0.95\columnwidth]{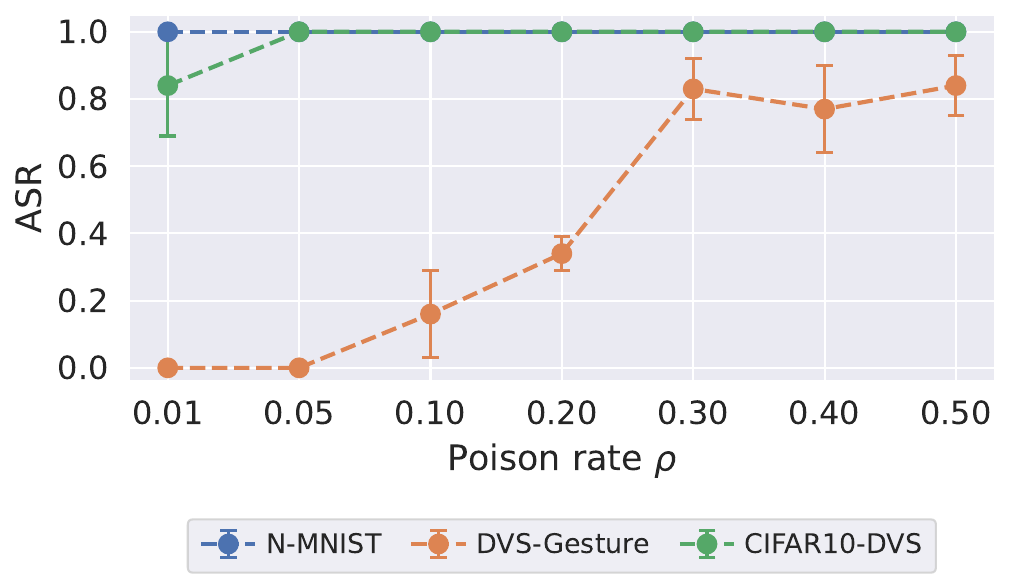}
  \caption{Poison-rate sweep on the convolutional victim using the concentrate trigger. Rates are expressed as fractions of the target-class pool. ASR saturates at the smallest tested rate on N-MNIST and from $\rho = 0.05$ on CIFAR10-DVS. On DVS-Gesture, whose target-class pool holds only $97$ samples, the ASR grows steadily with the poison rate but does not reliably reach $1.00$. Mean $\pm$ standard error over three seeds; Table~\ref{tab:poison-sweep} provides the exact values and sample counts.}
  \label{fig:poisonrate}
\end{figure}

\paragraph{Poison rate.}
We vary the poison rate to characterize the attack's sample requirements (Figure~\ref{fig:poisonrate} and Table~\ref{tab:poison-sweep}). On N-MNIST, concentrate reaches ASR $1.00$ at the smallest evaluated rate, $\rho=0.01$, corresponding to $59$ target-class samples and less than $0.1\%$ of the complete training set. On CIFAR10-DVS, the attack reaches $0.84$ with nine poisoned samples and $1.00$ with $45$ samples. Increasing the poison rate further does not improve ASR on either dataset.

DVS-Gesture behaves differently. Its target-class pool contains only $97$ samples, so although the ASR grows generally with the poison rate, it does not reach $1.00$ at any evaluated budget. These results show that poison fraction alone does not predict attack effectiveness. The absolute number of examples, dataset characteristics, victim fit, and seed variability all contribute to whether the temporal association is learned.

\begin{table}[b]
\centering
\caption{Attack success rate as a function of poison rate $\rho$, expressed as a fraction of the target-class pool. Mean $\pm$ SE over three seeds, with the actual number of poisoned samples in parentheses. Entries without an SE reached identical values across all seeds. The attack saturates at ASR $1.00$ from the smallest budget tested on N-MNIST and from $\rho=0.05$ on CIFAR10-DVS. DVS-Gesture, whose target-class pool holds only $97$ samples, never saturates, and its estimates carry wide error bars throughout.}
\label{tab:poison-sweep}
\begin{tabular}{lccc}
\toprule
 & N-MNIST & CIFAR10-DVS & DVS-Gesture \\
$\rho$ & (pool 5923) & (pool 900) & (pool 97) \\
\midrule
0.01 & $1.00 \pm 0.00$ \small(59)   & $0.84 \pm 0.15$ \small(9)   & $0.00 \pm 0.00$ \small(1) \\
0.05 & $1.00 \pm 0.00$ \small(296)           & $1.00 \pm 0.00$ \small(45)  & $0.00 \pm 0.00$ \small(4) \\
0.10 & $1.00 \pm 0.00$ \small(592)           & $1.00 \pm 0.00$ \small(90)           & $0.16 \pm 0.13$ \small(9) \\
0.20 & $1.00 \pm 0.00$ \small(1184)          & $1.00 \pm 0.00$ \small(180)          & $0.34 \pm 0.05$ \small(19) \\
0.30 & $1.00 \pm 0.00$ \small(1776)          & $1.00 \pm 0.00$ \small(270)          & $0.83 \pm 0.09$ \small(29) \\
0.40 & $1.00 \pm 0.00$ \small(2369)          & $1.00 \pm 0.00$ \small(360)          & $0.77 \pm 0.13$ \small(38) \\
0.50 & $1.00 \pm 0.00$ \small(2961)          & $1.00 \pm 0.00$ \small(450)          & $0.84 \pm 0.09$ \small(48) \\
\bottomrule
\end{tabular}
\end{table}

\begin{figure*}[t]
  \centering
  \includegraphics[width=0.95\textwidth, trim= 0pt 7pt 0pt 0pt, clip]{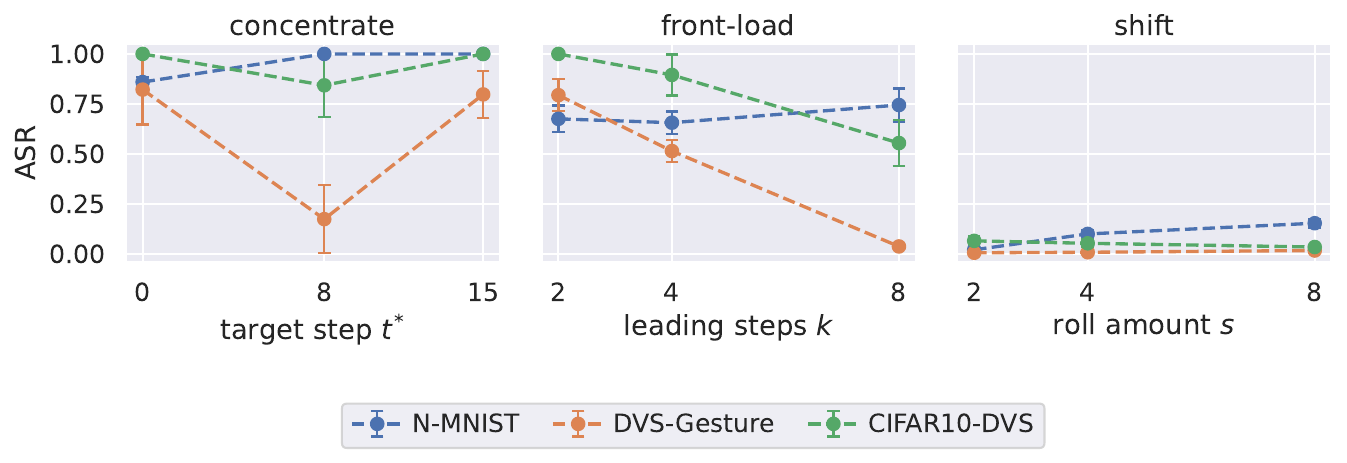}
  \caption{Trigger-shape sweep on the convolutional victim, reported as ASR for N-MNIST, DVS-Gesture, and CIFAR10-DVS. Each panel sweeps one trigger family over its shape parameter: concentrate over the target step $t^{*}$, front-load over the number of leading steps $k$, and shift over the roll amount $s$. On the two DVS datasets, concentrate is strongest at the ends of the interval and weakest in the middle, and front-load decays as its mass spreads over more steps; on the easier N-MNIST the attack stays strong across settings. Shift stays near zero throughout. Mean $\pm$ standard error over three seeds.}
  \label{fig:triggershape}
\end{figure*}

\paragraph{Trigger shape.}
Figure~\ref{fig:triggershape} examines how each trigger's temporal parameters affect ASR on the convolutional victims. Concentrate is highly sensitive to burst placement. On both DVS datasets, placing the burst near either boundary of the recording yields a higher ASR than placing it near the middle. On DVS-Gesture, for example, ASR is $0.82$ at $t^\ast=0$ and $0.80$ at $t^\ast=15$, compared with $0.18$ at $t^\ast=8$.

Front-load becomes weaker as its events are distributed across more leading time steps. This trend is consistent with the trigger becoming less distinct from the clean temporal profile as $k$ increases. In contrast, shift remains near zero for all evaluated roll amounts, indicating that changing the absolute temporal phase without strongly altering local event coincidence is insufficient to establish a reliable backdoor. Overall, the sweep shows that the shape and placement of the temporal redistribution are central to attack effectiveness.

\subsection{Why re-timing installs a backdoor}
\label{sec:mechanism}
The three transformations preserve the rate frame but produce attack success rates ranging from $1.00$ to nearly zero. This difference is consistent with the stateful spiking dynamics described in Section~\ref{sec:snn}. For a LIF neuron, the recurrence $H[\tau]=\beta V[\tau-1]+X[\tau]$ in \eqref{eq:lif-charge} makes the response depend on how the input is distributed across time. Events that arrive within the same or nearby bins can produce a different membrane trajectory from the same events spread across the complete recording, even when $\sum_\tau X[\tau]$ is unchanged.

Concentrate creates the strongest change in coincidence by packing a large fraction of the stream into a short temporal region. This produces a repeatable burst-like input pattern across poisoned target-class samples, which the optimizer can associate with the target label. Front-load produces a related but less abrupt change by moving event mass toward earlier time steps. Its decreasing ASR as $k$ increases is consistent with the transformed temporal profile becoming less concentrated and less distinct from clean data.

Shift provides a useful negative control. Except for events crossing the recording boundary, it preserves local inter-event intervals and primarily rotates the stream in time. Because the victim aggregates class evidence over all $T$ steps, this transformation does not consistently create a new coincidence pattern for the classifier to associate with the target class. This interpretation is consistent with its near-zero ASR across most configurations. The value of $0.22$ on the N-MNIST convolutional victim may result from boundary wraparound or architecture-specific sensitivity, but the current experiments do not isolate its precise cause.
 
\section{Defenses}
\label{sec:defenses}

We evaluate the attack against five established backdoor defenses and our own detector. Two of the established defenses can be adapted to a spiking victim in more than one way, so we run each of them in two variants. None of these defenses were designed for data with a time axis, so adapting each one requires a choice its original form did not have to make: where along the pipeline to collapse the $T$ time steps into one. As Section~\ref{sec:defense-results} shows, this choice largely decides whether a defense can see a timing trigger at all, so for each defense we describe both its original form and the change we made.

We consider that a defense is \emph{input-collapsed} if it inspects a sample only after the $T$ steps have been reduced to a rate frame, \emph{input-temporal} if it inspects the per-step structure of the input directly, and \emph{feature-collapsed} if it inspects a penultimate feature that has already been averaged over the $T$ steps. These labels describe our adaptations rather than the original methods.

\subsection{Defender Model}
\label{sec:defender-model}

All defenses operate under the clean-label data-poisoning threat model defined in Section~\ref{sec:threat}. The defender receives a victim trained on the poisoned set and, depending on the defense, must decide whether a backdoor is present, identify the poisoned samples, or remove the backdoor. We give the defender strong access: every model-based defense can read the victim's weights, and the clean holdout set it needs is drawn from the held-out test split rather than from the possibly poisoned training data. Neural Cleanse additionally receives input gradients, and pruning may modify the weights in place. This defender is stronger than a realistic one, which is deliberate, since a defense that fails under favorable conditions fails under realistic ones as well. Table~\ref{tab:defenses} summarizes what each defense inspects and where it stands relative to the time axis.

\begin{table}[b]
\centering
\footnotesize
\setlength{\tabcolsep}{4pt}
\caption{Defenses considered. The final column records where the $T$ time steps are collapsed, which is the adaptation choice that determines what each defense can perceive. The same labels are used in Figure~\ref{tab:detection}.}
\label{tab:defenses}
\begin{tabular}{@{}llll@{}}
\toprule
Defense & Access & Inspects & Time axis \\
\midrule
STRIP (rate)         & queries         & pred.\ entropy  & input-collapsed \\
STRIP (event)        & queries         & pred.\ entropy  & input-temporal \\
Act.\ Clustering     & weights         & penult.\ spikes & feature-collapsed \\
Spectral Sig.        & weights         & penult.\ spikes & feature-collapsed \\
Neural Cleanse (sp.) & weights, grads  & recovered mask  & constant over $T$ \\
Neural Cleanse (t.)  & weights, grads  & recovered gain  & per-step \\
Pruning              & weights (write) & firing rates    & feature-collapsed \\
\midrule
Temporal (ours)      & input only      & per-step mass   & input-temporal \\
\bottomrule
\end{tabular}
\end{table}

\subsection{Feature Extraction for Spiking Victims}
\label{sec:defense-features}

Activation Clustering, Spectral Signatures, and pruning all read a penultimate feature vector, which is clear in a normal network but not in a spiking one, where we first have to decide which quantity is the activation and how the $T$ steps are reduced to a single vector. We take the activation to be the spike output of the last spiking neuron before the classifier readout, not the membrane potential. For the convolutional victim this is the neuron after the first fully connected layer, giving $2048$ dimensions on N-MNIST and $512$ on the two DVS datasets; for SpikformerLite, abbreviated Spikformer in the tables, it is the neuron closing the final block's MLP, giving $128$ dimensions. We reduce over time by averaging, so the feature is the per-neuron firing rate across the $T$ steps, and for the transformer the average also runs over the $64$ tokens, since the classifier mean-pools them. These three widths, $2048$, $512$, and $128$, turn out to matter for what Spectral Signatures can recover.

Averaging over time is the natural choice: it produces exactly the fixed-length real-valued vector these defenses were built for, and it matches what the victim's own readout does before the logits. It is also the point at which a sample's temporal structure is discarded. We use it because it is what a defender following the published methods would do.

\subsection{Established Defenses}
\label{sec:defense-baselines}

STRIP~\cite{strip} detects triggered inputs at inference time by superimposing a candidate on many clean partner samples and measuring the entropy of the resulting predictions; a backdoored input keeps driving the model to the target class whatever it is blended with, so its prediction entropy stays unusually low. Superimposing on a $[T, 2, H, W]$ tensor can be defined in more than one way, so we use two variants. The event-level variant sums two tensors frame by frame, preserving the timing of both the candidate and the partner and therefore being input-temporal. The rate-frame variant collapses both inputs to their rate frames, sums them, divides by $T$, and re-expands the result into a stack that is constant across time, which throws away the timing of both and is therefore input-collapsed. Each candidate is blended with $100$ partners drawn from a pool of $200$ clean test samples, and we score $150$ non-target evaluation samples. Entropy is the Shannon entropy in nats of the softmax over the time-averaged logits, averaged across the $100$ blends, and the detection score is its negation so that low entropy means poison. The two variants do not give the victim the same amount of input, since the event-level blend is a plain sum while the rate-frame blend is divided by $T$, so they should be read as two separate detectors rather than a controlled comparison.

Activation Clustering~\cite{ac} assumes that poisoned and clean samples of the same class sit in different parts of feature space, and separates them by clustering the penultimate activations of a single class into two groups. We run it on the target class using the time-averaged spike features of Section~\ref{sec:defense-features}. Following the standard configuration, we reduce to $10$ independent components with FastICA, fall back to a PCA projection of the same rank if it fails to converge, and cluster with $2$-means. The smaller cluster is taken as the poison candidate, and the class is flagged when the silhouette score of the split is above $0.10$. Because the silhouette computation is quadratic in the sample count, we estimate it on a random subsample of at most $2000$ points.

Spectral Signatures~\cite{spectral} looks for poison along the main direction of variation in the feature covariance of a single class. We again use the target-class pool and the time-averaged spike features. We mean-center the feature matrix, take the top right singular vector by SVD, and score each sample by the squared size of its projection onto that vector. We report this continuous score directly, which is what lets a detection curve be drawn, and we also apply the removal rule of the original method, which discards the highest-scoring $\lceil 1.5\,\epsilon_{\mathrm{pool}} N \rceil$ samples. The rate $\epsilon_{\mathrm{pool}}$ is the poison rate of the pool the defense is given, which is not the attack's poison rate $\rho$; Section~\ref{sec:defense-pool} makes the difference explicit.

Neural Cleanse~\cite{neuralcleanse} reverse-engineers, for each candidate class, the smallest input change that forces the model to predict it, and flags a model when one class needs a far smaller change than the rest. Our spatial version optimizes a mask and a pattern of shape $[2, H, W]$, so the recovered trigger has a polarity axis but no time axis. It is applied to the rate frame and copied across the $T$ steps, so the search space contains only triggers that are constant in time. This follows the original method closely, and it is also why the spatial version cannot represent a timing trigger even in principle. Optimization uses Adam at learning rate $0.1$ for $1000$ steps on $512$ clean test samples, with an $L_1$ penalty of weight $10^{-2}$ on the mask. Because a constant-in-time search space cannot represent our trigger, we add a temporal version that searches the axis the spatial one cannot reach. It optimizes a per-step gain vector of length $T$, normalized by a softmax and rescaled to have mean one, applied element-wise over time, and penalized by its $L_1$ distance from a flat profile. This version recovers a burst-shaped gain instead of a spatial patch. It is a diagnostic rather than an exact recovery of our trigger, since a multiplicative gain does not keep each pixel's total event count fixed while our trigger does, and we report it as a cost rather than an anomaly index, since a single recovered profile per model does not support the across-class comparison the published index uses.

Two things about the anomaly index are worth noting. The published index is the median absolute deviation of the per-class mask norms, which becomes unstable when no class stands out and every mask looks the same, a situation this attack often produces. We therefore floor the deviation at $5\%$ of the median and clip the resulting index, and we note both as changes from the published statistic. We also report the index taken as the maximum over all classes, which is what a real defender can compute and what the flagging rule uses; an index measured at the known target class is available in our code but uses information a real defender would not have, so we do not report it.

Fine-Pruning~\cite{finepruning} removes a backdoor in two stages, first pruning the neurons that stay dormant on clean data, on the idea that a backdoor hides in that unused capacity, and then fine-tuning briefly to recover the clean accuracy that pruning costs. We implement the first stage only, and refer to the defense as pruning for that reason. Neurons in the penultimate spiking layer of Section~\ref{sec:defense-features} are ranked by their mean firing rate over $T$ on $2000$ clean holdout samples and pruned in ascending order, dormant neurons first, in batches of about $1\%$ of the layer width. Pruning a neuron zeroes its incoming weight and bias and the matching readout column, which silences the unit and removes its contribution to the logits, and it stops once clean accuracy has fallen $4$ points below its starting value. Any remaining attack success we report is therefore the backdoor surviving pruning alone, without a recovery epoch, and the comparison to the published two-stage defense should be read with that in mind.

\subsection{A Temporal-Axis Detector}
\label{sec:defense-temporal}

None of the defenses above looks at how events are distributed along the time axis, which is the only axis on which our trigger changes. The event-level STRIP variant comes closest, since it does not destroy the candidate's timing, but it still reduces the sample to a prediction entropy and never looks at the timing itself. A panel built only from the existing literature would therefore leave the attack untested on the one axis where it acts, and any claim of stealth would then be a result of the panel we chose rather than a real property of the attack. We build a temporal detector to close this gap and to act as the adaptive defender our threat model calls for.

The detector is model-free: it reads only the input tensor and never loads the victim, making it suitable as a data-cleaning filter before training. For each sample it forms the normalized per-step event mass $p[t]$ by counting events in each of the $T$ bins and dividing by the sample total, so that $p$ sums to one and does not depend on how many events a recording contains. From $p$ it computes five features. Three of them measure how concentrated the profile is: the normalized Shannon entropy, flipped so that larger values are more anomalous, the peak mass $\max_t p[t]$, and the Gini coefficient of the sorted profile. The fourth is the Kullback-Leibler divergence from $p$ to the mean profile of the clean training data, which captures changes in how the mass is spread, regardless of its shape. The fifth is a phase lag, found by cross-correlating the mean-centered profile against circularly shifted copies of the clean mean profile and taking how far the best-matching shift is from zero, scaled to the unit interval; it exists because the first four features do not react to a pure rotation of the time axis, which is exactly what the shift trigger does.

The decision boundary is set on clean data alone. We take the $99$th percentile of each feature across $2000$ clean training samples and score a sample by how far its most extreme feature exceeds its boundary. Because the boundary and standardization are set on clean data alone, no poisoned samples are used to set the detector's threshold, unlike Activation Clustering and Spectral Signatures, which have to estimate their statistics from the mixed pool they are tasked with filtering.
 
\subsection{Evaluation Pool}
\label{sec:defense-pool}

Activation Clustering and Spectral Signatures are training-set filters, so each is given a pool of target-class training samples in which the poisoned members must be found. We rebuild that pool from raw data rather than reusing stored tensors, applying the trigger on the fly to exactly the sample indices the attack poisoned.
Section~\ref{sec:results-augablation} tests directly whether this difference in augmentation, rather than the trigger, drives the one positive detection result in the panel.

One thing about the pool affects how the detection numbers are read. The clean side is capped at $3000$ samples while every poisoned sample is kept, so the pool's poison rate $\epsilon_{\mathrm{pool}}$ is not the attack's rate $\rho$. On N-MNIST at $\rho = 0.1$, the $592$ poisoned samples sit next to $3000$ clean ones, giving $\epsilon_{\mathrm{pool}} \approx 0.165$ rather than $0.1$, and both the detection rates and the removal budget of Spectral Signatures are computed against this pool rate.

\section{Defense Results}
\label{sec:defense-results}

Every number below is reported for the six victim cells, the convolutional SNN and SpikformerLite on each of the three datasets, at the attack configuration of Section~\ref{sec:evaluation}: target class $0$, three seeds, and the trigger placements fixed there. The attack is effective and invisible in the rate frame on all six cells; we now ask which defenses can still detect or remove it. Unless stated otherwise, every number is the mean over three seeds, with standard error, rounded to two decimal places.

\subsection{Detection Defenses}
\label{sec:results-detection}

\begin{figure*}[t]
  \centering
  \includegraphics[width=0.94\linewidth, trim= 11pt 0pt 30pt 0pt, clip]{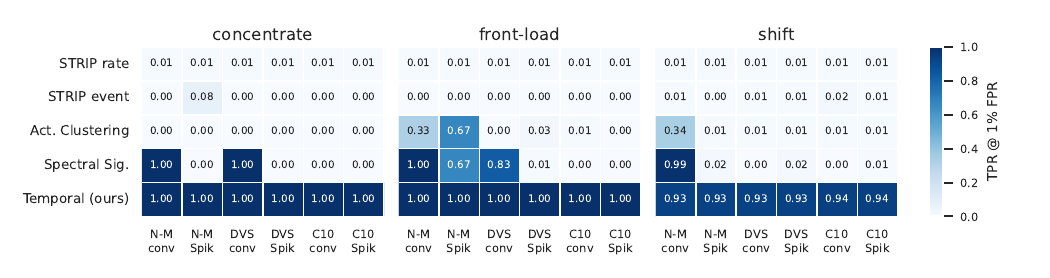}
  \caption{Detection performance (TPR at $1\%$ FPR) across the six victim cells, by trigger family. Rows are defenses; ours is the bottom row. Chance is $0.01$. DVS is DVS-Gesture, CIFAR is CIFAR10-DVS, Spik is SpikformerLite.}
  \label{tab:detection}
\end{figure*}

Figure~\ref{tab:detection} reports the true positive rate at a $1\%$ false positive rate for each detection defense, on each trigger family and each cell. By construction, the chance is $0.01$, so an entry at $0.01$ carries no information about the poison.

STRIP is ineffective in both variants. The rate-frame variant sits at the chance rate of $0.01$ across all eighteen entries, as expected: the trigger leaves the rate frame unchanged, so the rate-frame blend gives the victim a candidate that is identical to its clean version. The event-level variant keeps the candidate's timing but does no better, staying at or below $0.02$ everywhere except for a single $0.08 \pm 0.05$ on the N-MNIST Spikformer cell; mixing in a clean partner's events hides the timing change rather than showing it. Keeping the time axis in the blend is therefore necessary, but not enough to see the trigger.

Activation Clustering is uninformative. It sits at $0.00$ on concentrate in all six cells and at chance on shift in five of six. The three above-chance values all fall on front-load, and two of them ($0.33 \pm 0.33$ and $0.67 \pm 0.33$) have standard errors as large as their means, meaning a single seed carries the effect, so we do not count them as detections.

Spectral Signatures detects the poison in selected configurations. It reaches a TPR of $1.00$ on the N-MNIST and DVS-Gesture convolutional victims for concentrate, and $1.00$ and $0.83\pm0.13$, respectively, for front-load. In these cells, timestamp redistribution leaves a detectable trace in the time-averaged firing-rate features. It does not consistently separate poison in the remaining cells. Feature dimensionality, victim quality, architecture, and attack stability are possible explanations for this pattern, but the six evaluated cells do not support a causal conclusion.

The input-only temporal-mass detector reaches $1.00$ TPR for concentrate and front-load across all six cells and $0.93$--$0.94$ for shift. This shows that the evaluated transformations are visible when the detector directly measures their coarse temporal redistribution. Because its features were selected to capture concentration, distribution shift, and temporal phase, this result should not be interpreted as robustness against an adaptive attacker that preserves these statistics.

Overall, the results support two conclusions. First, input-collapsed adaptations cannot detect a trigger that is absent from the representation they inspect. Second, the evaluated time-averaged feature adaptations do not consistently detect the poison. These experiments do not establish that all temporally resolved feature-space or model-state defenses must fail.

\subsection{Neural Cleanse}
\label{sec:results-nc}

\begin{table}[b]
\centering
\footnotesize
\setlength{\tabcolsep}{5pt}
\caption{Neural Cleanse. The flag threshold on the anomaly index is $2$.  The last column shows the deviation-from-uniform cost of the temporal variant, which is reported only for contrast.}
\label{tab:nc}
\begin{tabular}{@{}llccc@{\;}c@{}}
\toprule
Dataset & Victim  & max anomaly & flags target & temporal-NC cost \\
\midrule
N-MNIST     & conv       & 0.47 & 0 / 3 & 1.19 \\
N-MNIST     & Spikformer & 2.90 & 2 / 3 & 1.55 \\
DVS-Gesture & conv       & 0.02 & 0 / 3 & 1.80 \\
DVS-Gesture & Spikformer & 1.63 & 0 / 3 & 1.79 \\
CIFAR10-DVS & conv       & 0.09 & 0 / 3 & 1.43 \\
CIFAR10-DVS & Spikformer & 0.61 & 0 / 3 & 1.49 \\
\bottomrule
\end{tabular}
\end{table}

Neural Cleanse in its spatial form is blind on all three convolutional cells (Table~\ref{tab:nc}), where the maximum anomaly index over classes never goes above $0.47$ against a flag threshold of $2$, and the true target is never the flagged class. This is what the method's search space predicts, since a constant mask across the $T$ steps cannot represent a change in event timing.

It is not blind everywhere. On the N-MNIST Spikformer cell, the maximum anomaly index averages $2.90$, and the true target class is flagged in two of three seeds, with per-seed values of $4.40$ and $3.36$. The DVS-Gesture Spikformer cell raises a false alarm on a non-target class in one seed. This does not mean the spatial mask has found the timing trigger, which it cannot. It means the poisoned transformer ended up with a target class that is unusually easy to reach with a small spatial change, which the anomaly index cannot distinguish from a real trigger.
 
The two model-based defenses, therefore, work on opposite cells. Spectral Signatures detects on the convolutional cells where Neural Cleanse is blind, and Neural Cleanse partly detects on the N-MNIST transformer cell where Spectral Signatures is at zero. Neither is reliable on its own, and on no cell do the two together give a defender a confident answer. The temporal variant, on the other hand, recovers a concentrated burst on every cell, with a deviation-from-uniform cost between $1.19$ and $1.80$, which shows that a reverse-engineering search can find the trigger once it looks along the time axis.

\subsection{Pruning}
\label{sec:results-pruning}

\begin{table}[b]
\centering
\footnotesize
\setlength{\tabcolsep}{6pt}
\caption{Pruning of the penultimate spiking layer in ascending clean firing rate, halted at a $4$-point clean-accuracy loss. This is the first stage of Fine-Pruning without the recovery fine-tuning stage. $\Delta$ values are drops from the pre-pruning value, expressed as fractions, so a large $\Delta$ASR would mean the backdoor was removed.}
\label{tab:pruning}
\begin{tabular}{@{}llcc@{}}
\toprule
Dataset & Victim & $\Delta$clean & $\Delta$ASR \\
\midrule
N-MNIST     & conv       & 0.04$\pm$0.00 & 0.00$\pm$0.00 \\
N-MNIST     & Spikformer & 0.04$\pm$0.00 & 0.00$\pm$0.00 \\
DVS-Gesture & conv       & 0.05$\pm$0.01 & 0.25$\pm$0.14 \\
DVS-Gesture & Spikformer & 0.05$\pm$0.00 & 0.00$\pm$0.00 \\
CIFAR10-DVS & conv       & 0.04$\pm$0.00 & 0.12$\pm$0.11 \\
CIFAR10-DVS & Spikformer & 0.04$\pm$0.00 & 0.28$\pm$0.27 \\
\bottomrule
\end{tabular}
\end{table}

Pruning does not remove the backdoor within the clean-accuracy budget it is given (Table~\ref{tab:pruning}). Attack success is unchanged in three cells and drops by at most $0.28$ on the others; the two largest drops, $0.25$ and $0.28$, have standard errors of $0.14$ and $0.27$, so they are seed effects rather than a consistent defensive gain. The reason is structural. A clean-label poison never touches the weights, so the trigger has to be learned using the same neurons that clean samples already use, and it ends up inside neurons that also do normal work. It does not create a dormant, unused neuron, so the assumption that pruning depends on it does not hold, and no tested pruning budget removes it. These numbers cover the pruning stage only; the recovery fine-tuning stage of Fine-Pruning is not applied, and since that stage retrains on clean data, it would be expected to restore clean accuracy rather than remove a backdoor that pruning failed to isolate.

\subsection{Is the Spectral Signal the Trigger or an Augmentation Gap?}
\label{sec:results-augablation}

The poisoned samples enter the filter pool without augmentation while the clean samples are augmented, so one might ask whether Spectral Signatures separates poison from clean based on augmentation rather than on the trigger. Table~\ref{tab:augablation} tests this on the frozen victim, comparing the reported case against one where the poison is replaced by augmented clean samples, and one where augmentation is applied to both sides. We do not interpret any cell that does not first reproduce the reported separation, since the question only makes sense where there is a separation to explain. On the two cells where Spectral Signatures actually detects, we can rule out the confound: augmentation alone does not separate the samples (AUC $0.50$ and $0.48$, against a chance value of $0.50$), and the separation stays at $1.00$ when both sides are augmented, so the detection comes from the re-timing itself, and the result in Figure~\ref{tab:detection} holds. In the other four cells, the poison case does not reproduce a separation, so we mark them inconclusive rather than negative.

\begin{table}[b]
\centering
\footnotesize
\setlength{\tabcolsep}{5pt}
\caption{Augmentation-confound ablation for Spectral Signatures, reported as AUC on the frozen victim. Chance is $0.50$. Cells where the poison condition falls below $0.90$ are marked inconclusive, because Spectral Signatures does not separate the poison there in the first place, which makes the confound question moot rather than answered.}
\label{tab:augablation}
\begin{tabular}{@{}llcccl@{}}
\toprule
 & & clean vs & clean vs & aug.\ both & \\
Dataset & Victim & poison & aug.\ clean & sides & Verdict \\
\midrule
N-MNIST     & conv       & 1.00 & 0.50 & 1.00 & confound rej. \\
N-MNIST     & Spikformer & 0.83 & 0.49 & 0.80 & inconclusive \\
DVS-Gesture & conv       & 1.00 & 0.48 & 1.00 & confound rej. \\
DVS-Gesture & Spikformer & 0.26 & 0.50 & 0.33 & inconclusive \\
CIFAR10-DVS & conv       & 0.66 & 0.53 & 0.70 & inconclusive \\
CIFAR10-DVS & Spikformer & 0.65 & 0.50 & 0.61 & inconclusive \\
\bottomrule
\end{tabular}
\end{table}



\section{Related Work}
\label{sec:related_work}

\subsection{Spiking Neural Networks}

SNNs replace the continuous activations of DNNs with spiking neurons that communicate through discrete events over time, integrating incoming spikes into a membrane potential and firing once a threshold is reached~\cite{gerstner2002spiking,tavanaei2019deep}. The main obstacle to training these networks is the non-differentiability of the spike function, which prevents the direct use of gradient-based optimization~\cite{neftci2019surrogate}. For direct SNN training, two main lines of work address this: Spike-timing-dependent plasticity (STDP) adjusts synaptic weights according to the relative timing of pre- and post-synaptic spikes~\cite{kheradpisheh2018stdp}, and surrogate gradient methods, which approximate the spike function with a differentiable surrogate during the backward pass, enabling standard backpropagation~\cite{neftci2019surrogate}. On the other hand, another approach consists of converting a pretrained DNN into an equivalent SNN, at the cost of longer inference and higher power consumption due to an increased number of activations~\cite{ding2021optimal}.

\subsection{Backdoor Attacks}
Backdoor attacks were first demonstrated on DNNs by Gu et al. with BadNets, which injects a fixed patch trigger into a subset of the training data and relabels those samples to a target class~\cite{gu2019badnets}. This dirty-label formulation is effective, but obvious under label-based inspection, since the label of a poisoned sample no longer matches its contents. Clean-label attacks were introduced to remove this weakness. Turner et al. inject poisoned samples that remain visually plausible and label-consistent but are perturbed, making them hard to cleanly classify, so the model learns to rely on the easier-to-learn trigger feature~\cite{turner2019label}. Clean-label backdoor attacks have since been extended beyond static images, for instance, to video recognition models~\cite{zhao2020cleanlabel} and to networks trained from scratch via gradient matching in Sleeper Agent~\cite{souri2022sleeper}.

Conversely, to the best of our knowledge, backdoor attacks on SNNs have so far focused on dirty-label poisoning. The first backdoor attack on spiking models and neuromorphic datasets, presented by Abad et al., used static and moving square triggers that appear in all frames of the poisoned event streams while assigning them a chosen target label~\cite{abad2022poster}. Sneaky Spikes extends this line by exploring a richer family of triggers within neuromorphic data, allowing their position and polarity to vary over space and time and achieving high attack success rates and stealthiness with limited impact on clean accuracy~\cite{abad2024sneaky}. Follow-up work has moved these triggers into the physical world: Flashy Backdoor demonstrates real-world environment backdoor attacks on SNNs with DVS cameras by reproducing flash-like triggers through controllable light sources, again using relabeled poisoned samples during training~\cite{physicalsnn}.
Other studies focus on distributed settings. Abad et al.~\cite{abad2024timedistributed} were among the first to study backdoor attacks on SNNs in a federated learning setting, distributing trigger activity over time and across malicious clients. Spikewhisper took another approach by proposing temporal spike backdoor attacks on federated neuromorphic learning, where different malicious clients insert local triggers in different time slices of the neuromorphic sample~\cite{fu2024spikewhisper}.
In both cases, triggered samples are trained with the target label to implant the backdoor. Jin et al.~\cite{jin2024poisoning} introduce a generic data-poisoning-based backdoor framework against supervised SNN learning rules and analyze backdoor robustness across different training rules and between SNNs and ANNs. All these SNN backdoor attacks adopt the standard dirty-label assumption that poisoned inputs containing the trigger are relabeled to the target class during training. In contrast, the clean-label setting has not been examined for SNNs. Our work fills this gap by introducing a clean-label backdoor whose trigger is encoded in the timing of neuromorphic events.

\section{Conclusions \& Future Work}
\label{sec:conclusions}

We presented the first clean-label backdoor attack on spiking neural networks. Instead of a spatial pattern, the trigger is a fixed change to the event timestamps that keeps the number of events at each pixel (and therefore the rate frame) the same for every sample, so the poison is invisible to any check that collapses the time axis ($\mathrm{SSIM}=1.0$, $L_0=L_\infty=0$), while still changing how the events are spread over time, which is what the network reacts to. Across three neuromorphic datasets and both convolutional and transformer SNNs, a concentrate trigger achieves an ASR of up to $1.00$ from only a few dozen retimed target-class samples, with clean accuracy preserved in most configurations.
 
Testing the attack against a set of established defenses adapted to spiking victims shows that the standard set of defenses is insufficient for neuromorphic data. Every defense that reduces a sample to its rate frame before looking at it is blind by design, and even the feature-space filters find the poison only in selected configurations; identifying the roles of architecture, model quality, and representation dimensionality requires further study. Detecting it reliably instead needs a measure computed directly on how the events are spread over time, which our temporal detector provides. Several directions are left open for future work. Relaxing the requirement that the rate frame stay unchanged would trade some stealth for a stronger attack on harder temporal tasks; a timing trigger might also be reproduced physically through a DVS sensor, as has already been shown for spatial triggers~\cite{physicalsnn}; and the gap we find points to the need for time-aware defenses built for neuromorphic data instead of adapted from the frame-based setting.

\bibliographystyle{ACM-Reference-Format}
\bibliography{bib}


\appendix

\section{Artifact \& Reproducibility}
All code, configurations, and the exact random seeds needed to reproduce every table and figure in this paper will be available in our repository, for academic use only. The implementation uses PyTorch with the SpikingJelly framework~\cite{spikingjelly}; exact package versions and setup instructions are in the repository, and all experiments were run on a single NVIDIA H200 GPU running SUSE Linux, with PyTorch 2.8. The model, training, and trigger settings are provided in Section~\ref{sec:evaluation}, and the defense settings in Section~\ref{sec:defenses}.

\section*{Use of Generative AI}
The authors used generative AI tools to assist with grammar refinement. All scientific content, including the content of the tables and algorithms, was developed and verified solely by the authors.

\section*{Open Science}

In line with the open-science goals of the community, all artifacts underpinning this paper, the attack implementation, the victim models, every defense we evaluate, the trigger and defense configurations, and the scripts that produce each table and figure, are made publicly available. For double-blind review they are provided through an anonymized repository; upon acceptance we will release them in a permanent, publicly archived repository with a citable DOI. The datasets used (N-MNIST, DVS-Gesture, and CIFAR10-DVS) are public benchmarks obtained as described in the reproducibility appendix.

\end{document}